\newif\ifanonymous
\anonymousfalse   

\documentclass[10pt,conference]{IEEEtran}
\IEEEoverridecommandlockouts

\usepackage{cite}
\usepackage{amsmath,amssymb,amsfonts}
\usepackage{graphicx}
\usepackage{textcomp}
\usepackage{xcolor}
\usepackage{url}
\usepackage{hyperref}
\usepackage{booktabs}
\usepackage{listings}
\usepackage{tikz}
\usetikzlibrary{positioning,fit,arrows.meta,backgrounds,calc}

\definecolor{reqblue}{RGB}{219,234,247}
\definecolor{reqblueline}{RGB}{88,133,175}
\definecolor{stdlav}{RGB}{233,226,245}
\definecolor{stdlavline}{RGB}{136,113,181}
\definecolor{metgreen}{RGB}{222,240,223}
\definecolor{metgreenline}{RGB}{97,153,102}
\definecolor{tstamber}{RGB}{252,238,215}
\definecolor{tstamberline}{RGB}{193,142,62}
\definecolor{outgray}{RGB}{238,238,238}
\definecolor{outgrayline}{RGB}{120,120,120}
\definecolor{spinecol}{RGB}{170,60,60}

\definecolor{codegray}{RGB}{245,245,245}
\definecolor{codekw}{RGB}{0,0,160}
\definecolor{codestr}{RGB}{140,40,40}
\definecolor{codecmt}{RGB}{80,120,80}
\lstdefinelanguage{JavaScript}{
  morekeywords={const,let,var,function,return,module,exports,require,import,from,export,default,test,expect,async,await,new},
  sensitive=true,
  morecomment=[l]{//},
  morecomment=[s]{/*}{*/},
  morestring=[b]',
  morestring=[b]",
  morestring=[b]`
}

\def\BibTeX{{\rm B\kern-.05em{\sc i\kern-.025em b}\kern-.08em T\kern-.1667em\lower.7ex\hbox{E}\kern-.125emX}}

\begin{document}

\title{Integrating High-Level Requirements to Low-Level Tests with Machine-Readable V\&V Specifications\\
}

\author{
\IEEEauthorblockN{Mansur Arief}
\IEEEauthorblockA{\textit{Industrial and Systems Engineering,} \\
\textit{King Fahd University of Petroleum}\\
\textit{and Minerals (KFUPM)}, Saudi Arabia\\
mansur.arief@kfupm.edu.sa}%
\\
\IEEEauthorblockN{Ali Akarma}
\IEEEauthorblockA{
\textit{Faculty of Computer and Information Systems} \\
\textit{Islamic University of Madinah} \\
Madinah, Saudi Arabia \\
443059463@stu.iu.edu.sa
}
\\
\and
\IEEEauthorblockN{Nur Ahmad Khatim}
\IEEEauthorblockA{\textit{Division of Information Science,} \\
\textit{Nara Institute of Science}\\
\textit{and Technology (NAIST), Japan}\\
nurahmadkhatim@gmail.com}%
\\
\IEEEauthorblockN{Ahmad Alfan Alfian Irfan}
\IEEEauthorblockA{\textit{Information Technology} \\
\textit{Universitas Muhammadiyah Yogyakarta}\\
Yogyakarta, Indonesia\\
ahmad.alfan.ft23@mail.umy.ac.id}}%

\maketitle

\begin{abstract}
Modern software teams have mature tools for low-level testing, such as pytest, JUnit, and Jest, which make it inexpensive to write unit tests and run them on every commit. Systems engineering, in parallel, has developed rigorous principles for design verification and validation (V\&V), which has worked very well across engineering discipline to align user expecations and requirements with developers' deliverables. In practice, however, the two rarely connect, and the link between users' high-level requirements and the low-level tests that machines actually run is maintained by hand, if at all. This gap is increasingly costly for AI-enabled and cyber-physical systems, for which regulators now ask for traceable evidence that high-level requirements are met, while raw test results provide little of the structure such evidence requires. We introduce \textsc{VNVSpec}, an open-source framework that makes V\&V specifications machine-readable and executable. With this framework, users state high-level requirements directly or import them from catalogs derived from published standards. Then, the framework checks requirement quality, supports decomposition into module-level requirements with explicit metrics and acceptance criteria, links these requirements to test results through a traceability graph, and compiles the collected evidence into verdicts and audit-ready reports. We evaluate the framework by self-application, in which it is continuously assessed in CI against its own specification of 36 requirements verified by 449 tests, completed within limited time which scales linearly and thus can handle up to 10,000 requirements. We also discuss how the framework extends to testing black-box AI models and AI coding agents. The framework, its full test suite, the catalogs, and the benchmark scripts are available at \url{https://github.com/ai-vnv/vnvspec}.

\end{abstract}


\section{Introduction}
\label{sec:intro}

Software engineering and systems engineering have each solved one half of the same verification and validation (V\&V) problem. On the software side, low-level testing has become inexpensive: a developer writes a unit test in pytest \cite{pytest} or a property-based test in Hypothesis \cite{maciver2019hypothesis}, the continuous integration (CI) system runs it on every commit, and a pass-or-fail result comes back in minutes. On the systems side, high-level V\&V has become rigorous: the INCOSE handbook \cite{incose2023handbook}, ISO/IEC/IEEE 29148 \cite{iso29148}, and the NASA Systems Engineering Handbook \cite{nasa2016se} define how to write verifiable requirements, how to decompose them, how to plan verification activities, and how to maintain traceability from stakeholder need down to test evidence. However, both fields are often disconnected, which makes it difficult to trace and answer basic questions about the relationship between high-level requirements and low-level tests.

Requirements traceability has been studied for a few decades \cite{gotel1994analysis,cleland2014software}, and model-based systems engineering (MBSE) tools built around SysML \cite{friedenthal2014practical,madni2018mbse} do maintain these links. However, MBSE toolchains are heavyweight, live outside the developer workflow, and rarely connect to the test frameworks that developers actually run. In practice, high-level requirements live in a procurement or tender document or a ticket system, unit tests live in the code repository, and the mapping between the two is reconstructed manually by the technical team, if it is maintained at all. As a result, engineers can rarely answer, at any given commit, three basic questions: which user requirements are covered by which tests, which requirements have no test at all, and which standard clauses the current test evidence actually supports.

The rise of AI applications makes this gap more consequential, for at least three reasons. First, AI-enabled systems are hard to specify, and their requirements practices lag behind their testing needs \cite{amershi2019software,vogelsang2019requirements,rahimi2019toward}. Second, regulation has arrived: the EU AI Act \cite{euaiact} requires technical documentation with exactly this requirements-to-evidence structure, and standards such as ISO/PAS 8800 \cite{isopas8800}, ISO 21448 \cite{iso21448}, and UL 4600 \cite{ul4600} ask for argued, traceable safety evidence for AI and autonomous systems. Third, AI coding agents now write a growing share of production code. An agent that generates code from a user request has no standard way to record what the user asked for, how that request was decomposed, and which tests demonstrate that each part was met; without such a record, the output is unauditable by construction.

\begin{figure*}[t]
  \centering
  \includegraphics[width=\linewidth]{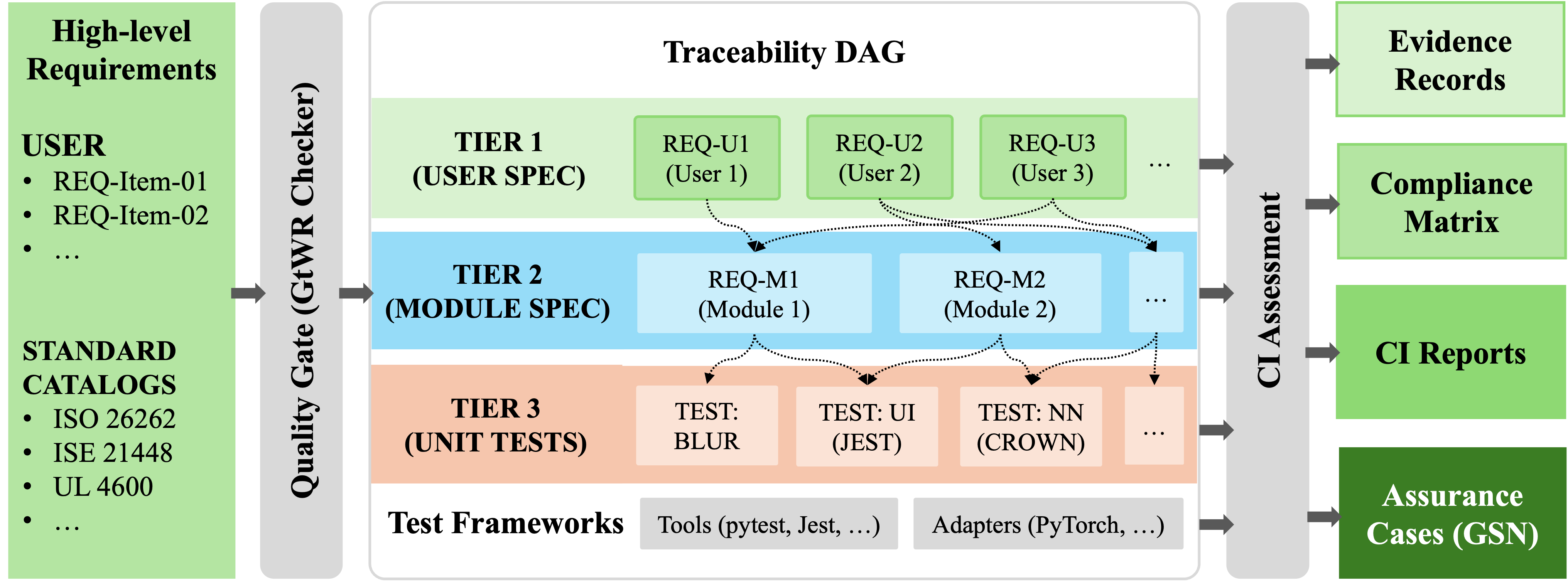}
  \caption{The \textsc{VNVSpec} framework. High-level requirements enter from users or from standards-derived catalogs and pass a quality gate. They are decomposed into module-level requirements with explicit metrics and further into test cases whose results are captured as evidence records by existing test tools. Every link is an edge in a traceability DAG. Assessment rolls evidence up the graph into verdicts and produces role-specific outputs: compliance matrices for auditors, CI reports for developers, and GSN assurance cases for safety engineers.}
  \label{fig:framework}
\end{figure*}

We address this gap by introducing \textsc{VNVSpec}, an open-source framework that makes the V\&V specification itself a typed, machine-readable, and executable artifact, which is summarized in Fig.~\ref{fig:framework}. The idea is to treat the specification as data: requirements, interface contracts, operational design domains, hazards, and evidence records are all typed objects that can be validated, composed, diffed, and traced. High-level requirements enter either from the user directly or from catalogs of curated requirements derived from published standards and framework documentation. The framework then supports the decomposition path that systems engineering prescribes, from high-level requirement, to module-level requirement with a defined metric and acceptance criterion, to concrete test cases. Test results flow back in from existing tools, and the framework rolls them up through the traceability graph into verdicts, compliance matrices, and assurance-case exports.

Our contributions are two-fold. \textbf{The first} is a \emph{typed specification language} for V\&V, with immutable, serializable models for requirements, contracts, hazards, operational design domains, and evidence, together with a quality checker that flags vague or unverifiable requirements at authoring time. \textbf{The second} is a \emph{bridge to existing test tooling}: a pytest plugin, JUnit XML ingestion that turns results from JavaScript, Java, and C\texttt{++} test runners into requirement-linked evidence, an evidence collector for analysis scripts and formal verification runs, and model adapters that assess PyTorch and HuggingFace models directly. We illustrate when each evidence route applies through three worked examples, using pytest, Jest, and CROWN-based certified output bounds. We evaluate the framework by self-application and by measurement: the experiments in Section~\ref{sec:eval} quantify the runtime of every stage of the pipeline, its scaling behavior up to 10,000 requirements, the overhead the integrations impose on existing test runs, and the detection behavior of the quality checker under seeded defects. We further envision agentic modules that decompose high-level specifications into unit tests, which motivates our future work.

The remainder of this paper is organized as follows. Section~\ref{sec:background} reviews related work on requirements quality, traceability, executable specifications, MBSE, and AI standards. Section~\ref{sec:framework} describes the design of the framework. Section~\ref{sec:workflow} illustrates the workflow, the choice among evidence routes, and the experimental evaluation. Section~\ref{sec:disc} discusses the experimental results in light of the claimed contributions. Section~\ref{sec:discussion} presents limitations and future work, and Section~\ref{sec:conclusion} concludes.

\section{Related Work}
\label{sec:background}

\subsection{Requirements Quality and Traceability}

Requirements engineering doctrine is well established. ISO/IEC/IEEE 29148 \cite{iso29148} defines the characteristics of good requirements. The INCOSE Guide to Writing Requirements \cite{incose2023gtwr} turns these characteristics into concrete, checkable rules: use active voice, avoid vague terms, state a single verifiable condition, and so on. Traceability, the ability to follow a requirement's life in both directions, was formalized by Gotel and Finkelstein \cite{gotel1994analysis}, and its practical failures are well documented \cite{cleland2014software}. A substantial line of work treats lost traceability as a recovery problem and reconstructs links after the fact with word embeddings and sequential semantics \cite{zhao2017improved,chen2019enhancing}. Recovery is needed precisely because links were never captured when the work was done: traceability maintained by hand decays, and the field has repeatedly called for traceability that arises as a by-product of normal work rather than as a separate activity. \textsc{VNVSpec} takes this position literally: trace links are typed objects created where the work happens, in code and in CI, and a scanner derives links automatically from requirement-ID references in the codebase.

\subsection{Model-Based Systems Engineering}

MBSE replaces document-centric systems engineering with models, typically in SysML \cite{friedenthal2014practical}. MBSE tools do maintain requirement decomposition and verification links, and the approach is standard practice in aerospace and automotive programs \cite{madni2018mbse}. Our aim is not to replace MBSE. The observation is that most software teams, including most teams building AI systems, will not adopt a SysML toolchain, and MBSE tools do not integrate with pytest or a JavaScript test runner. \textsc{VNVSpec} carries the useful discipline of MBSE (typed requirements, decomposition, and traceability) into the developer's own environment: plain Python objects, YAML or TOML files in the repository, and CI.

\subsection{Executable Specifications and Requirements-as-Code}

Closer to our approach are practices that keep requirements next to the code. Behavior-driven development (BDD) expresses acceptance criteria in the Gherkin language and binds them to executable step definitions in tools such as Cucumber \cite{cucumber}. Requirements-as-code tools such as Doorstop \cite{doorstop2014} and Sphinx-needs \cite{sphinxneeds} store requirement items as version-controlled text with typed links, and Eclipse Capra \cite{maro2016capra} maintains traceability links across heterogeneous development artifacts. These tools share our premise that requirements belong in the repository, but each covers only a slice of the requirement-to-evidence chain: BDD binds acceptance scenarios to tests but has no requirement quality checking, standards mapping, or verdict semantics; Doorstop and Sphinx-needs manage requirement items and links but do not consume test results as evidence; and Capra manages links without a specification model or assessment logic. \textsc{VNVSpec} combines these slices in a single typed model and adds the layers an audit requires: quality gates, standards registries, conservative verdict roll-up, and assurance-case export.

\subsection{Testing of ML-Enabled Systems and AI Standards}

Software engineering for machine learning is an active field. Amershi et al.\ documented how ML changes the development workflow \cite{amershi2019software}, and Sculley et al.\ described the resulting technical debt \cite{sculley2015hidden}. Testing of ML programs has been on the forefront research agenda in safety, security, and reliability domains since at least Nakajima and Bui's work on dataset coverage as a test adequacy criterion \cite{nakajima2016dataset}, if not earlier. The ML Test Score \cite{breck2017ml} offers a rubric of tests that production ML systems should have, and Zhang et al.\ survey the ML testing landscape \cite{zhang2022machine}. On the documentation side, model cards \cite{mitchell2019model} and datasheets \cite{gebru2021datasheets} record properties of models and datasets. These lines of work provide content (what to test and what to document) but not structure: none of them defines a machine-checkable chain from a stated requirement through a metric to a test result. Formal neural-network verification \cite{katz2017reluplex, xu2020autolirpa} provides strong evidence for specific properties; in our framing they are evidence generators that need a formal specification framework.

Finally, a cluster of standards now targets AI and autonomous systems directly: ISO/PAS 8800 for road-vehicle AI safety \cite{isopas8800}, ISO 21448 for safety of the intended functionality \cite{iso21448}, UL 4600 for autonomous products \cite{ul4600}, the NIST AI Risk Management Framework \cite{nistairmf}, and the EU AI Act \cite{euaiact}, whose Annex IV specifies required technical documentation. All of them presuppose the same underlying structure: identified hazards, derived requirements, defined metrics, and traceable evidence. Assurance cases in Goal Structuring Notation (GSN) \cite{kelly2004goal,gsn2021standard} are the established notation for arguing that this structure is complete. Today, teams typically assemble this material manually for each audit, which is cumbersome and error-prone. Our framework addresses these needs.

\section{The Proposed Framework}
\label{sec:framework}

Fig.~\ref{fig:framework} shows the overall framework, whose design follows four principles. First, \emph{specifications are data}: every concept is a typed, immutable object that serializes losslessly to JSON, YAML, and TOML, so a specification is a reviewable, diffable file in the repository rather than a document in a separate tool. Second, the framework \emph{meets developers where they are}: it does not run tests itself, but consumes the output of the test tools teams already use and contributes the layer those tools lack, the mapping from test results back to requirements. Third, \emph{standards are data too}: clauses and curated best practices ship as importable registries and catalogs. Fourth, \emph{traceability is an enforced DAG}: every link is an explicit edge in a directed acyclic graph (i.e., a graph whose links never loop back on themselves), and cyclic links are rejected at construction time, so rolling coverage up the graph is always well defined. We discuss each component briefly.

\subsection{The Specification Language}
\label{sec:lang}

A \texttt{Requirement} carries an identifier, a normative statement, a rationale, a verification method (test, analysis, inspection, demonstration, or formal proof), acceptance criteria, a priority, and optional links to source documents and standards clauses. A \texttt{Spec} aggregates requirements, hazards, an operational design domain, and input--output contracts, and validates global properties at construction time, such as uniqueness of requirement and hazard IDs. Specs can be written in Python, YAML, or TOML and loaded with \texttt{Spec.from\_file()}; the three formats round-trip without loss. This matters in practice because requirement authors and test authors are often different people with different preferred tools.

The three tiers in Fig.~\ref{fig:framework} are conventions we adopt. A tier-1 requirement states what the user needs. A tier-2 requirement refines it with a technical metric and a threshold, and \texttt{derives\_from} links record the decomposition. A tier-3 artifact is a test whose result becomes an \texttt{Evidence} record linked by a \texttt{verifies} edge. Systems engineering handbooks describe exactly this decomposition \cite{incose2023handbook,nasa2016se}; the framework makes each step a checkable object with actual tests.

\subsection{Requirement Quality Checking}
\label{sec:gtwr}

Vague requirements are the earliest and cheapest defect to catch. \texttt{Requirement.check\_quality()} implements eight rules derived from the systems engineering guidelines \cite{incose2023gtwr}: detection of vague and unbounded terms (``fast,'' ``user-friendly,'' ``as appropriate''), of escape clauses, of multiple conditions packed into one statement, of missing verification criteria, of passive voice that hides the responsible actor, and related defects. Profiles adjust strictness by context, since a formal safety requirement and a web-app requirement should not be held to identical phrasing rules. The checker runs at authoring time and in CI, so a requirement that cannot be verified is flagged before anyone writes a test against it.

\subsection{Standards Registries and Requirement Catalogs}
\label{sec:catalogs}

\textsc{VNVSpec} provides two kinds of reusable content: registries and catalogs. \textit{Registries} are clause databases for published standards such as ISO/PAS 8800, ISO 21448, UL 4600, the EU AI Act, and NIST AI RMF. Requirements reference clauses by identifier, and the compliance exporter (Section~\ref{sec:export}) computes, for any chosen standard, which clauses are addressed by which requirements.

\textit{Catalogs} are curated requirement sets that encode published best practices for specific frameworks. Each catalog requirement passes a six-criteria inclusion gate: it must be actionable, verifiable, sourced from authoritative documentation, version-pinned to the target framework, mapped to standards clauses where applicable, and covered by an automated compatibility test. Currently,  five catalogs hold 116 requirements, of which 71\% carry at least one standards mapping (Table~\ref{tab:catalog}). A team starting a FastAPI service does not begin from a blank page; it begins from \texttt{Spec.extend()} over the security, observability, and API-design catalogs, then adds project-specific requirements. This is the ``standards as high-level requirements'' entry path in Fig.~\ref{fig:framework}: the user can state requirements directly, or state which catalogs and standards apply.

\begin{table}[t]
\caption{Example requirement catalogs. ``Mapped'' counts requirements carrying at least one standards-clause mapping.}
\label{tab:catalog}
\centering
\footnotesize
\setlength{\tabcolsep}{2.5pt}
\resizebox{\columnwidth}{!}{%
\begin{tabular}{@{}llcc@{}}
\toprule
\textbf{Catalog} & \textbf{Primary sources} & \textbf{Reqs.} & \textbf{Mapped} \\
\midrule
\texttt{ml.pytorch\_training} & PyTorch docs, NIST AI RMF & 32 & 20 \\
\texttt{ml.huggingface\_inference} & Transformers docs & 25 & 14 \\
\texttt{web.fastapi} & OWASP API Top 10 \cite{owasp2023api} & 22 & 20 \\
\texttt{web.sqlalchemy} & SQLAlchemy 2.0, Alembic docs & 18 & 9 \\
\texttt{optimization.pyomo} & Pyomo docs & 19 & 19 \\
\midrule
\multicolumn{2}{@{}l}{Total} & 116 & 82 \\
\bottomrule
\end{tabular}}
\end{table}

\subsection{Traceability}
\label{sec:trace}

A \texttt{TraceLink} is a typed edge: source, target, and a relation such as \texttt{derives\_from}, \texttt{verifies}, \texttt{mitigates}, or \texttt{maps\_to}. \texttt{build\_trace\_graph()} assembles links into a directed graph and rejects cycles, so the requirement-to-evidence structure is a DAG by construction. Because hand-maintained links decay, \texttt{auto\_trace()} scans the project tree for requirement-ID references, in test names, markers, docstrings, and comments, and generates \texttt{verifies} links automatically. The practical effect is that a developer who writes \texttt{@pytest.mark.vnvspec("REQ-001")} above a test has already done the traceability work.

\subsection{Evidence Collection and Test Integration}
\label{sec:evidence}

Evidence records are the point where the specification meets actual test results. An \texttt{Evidence} record captures a verification activity: which requirement it verifies, the kind of activity (test, analysis, inspection, demonstration, simulation, or formal proof), a verdict of pass, fail, or inconclusive, a timestamp, an artifact URI, and structured details such as metric values or certified bounds. Four routes produce evidence:

\begin{itemize}
  \item The \texttt{pytest-vnvspec} plugin captures test outcomes during a normal pytest run and emits evidence records for every test marked with a requirement ID. It validates marker references against the spec at collection time, generates inconclusive evidence for any test-method requirement with no linked test, and supports a configurable fail-on policy (\texttt{any}, \texttt{blocking}, or \texttt{never}).
  \item The JUnit XML ingester converts result files from any runner that emits the format, which covers the JavaScript, Java, and C\texttt{++} ecosystems, into evidence. The low-level tool does not need to know \textsc{VNVSpec} exists.
  \item \texttt{EvidenceCollector}, a context manager, wraps ad-hoc verification scripts, such as an analysis notebook, a black-box validation run, or a formal verification tool, and turns assertions and explicit verdicts into evidence records, validating every requirement against the spec.
  \item Adapters (such as \texttt{TorchAdapter} for \texttt{nn.Module} models for PyTorch, \texttt{TransformerAdapter}, \texttt{AutoregressiveAdapter}, and \texttt{VLMAdapter} for HuggingFace encoder, generative, and vision-language models) wrap a model behind a common protocol and drive it over assessment data with budget-aware batching, out-of-memory recovery, and activation and gradient hooks, emitting per-requirement evidence directly.
\end{itemize}

Assessment rolls the collected evidence up the trace graph into a report. The verdict logic is deliberately conservative: under the default strict policy, a requirement with any inconclusive evidence and no failures is judged inconclusive, reflecting the principle that the absence of evidence should never be mistaken for compliance. The CLI exposes the outcome through structured exit codes (0 for pass, 1 for failures, 2 for inconclusive, 3 for a spec validation error, 4 for a usage error, and 5 for an internal error), so a CI pipeline can gate on each condition distinctly.

\subsection{Reporting and Export}
\label{sec:export}

A single specification and a single body of evidence serve several audiences. The exporters produce an HTML report and Markdown summary for developers; a static V\&V dashboard site with per-requirement detail pages, a standards-compliance table, and a version-history timeline; an XLSX/CSV compliance matrix with per-clause gap analysis for auditors; a GSN assurance case in Mermaid notation \cite{gsn2021standard} for safety engineers; and an EU AI Act Annex IV technical-documentation skeleton populated from the spec for regulatory teams. A locally generated badge SVG and a Shields.io endpoint JSON surface the current verdict on the repository README. A GitHub Actions integration runs assessment on every push, publishes the badge, comments verdict summaries on pull requests, and diffs reports between commits so that a silently vanished piece of evidence is caught as a regression.

\section{Workflow, Routes, and Evaluation}
\label{sec:workflow}

\subsection{Workflow Example}
We illustrate the intended use via example: a team building an image classification service. The team imports the PyTorch training catalog, adds a user-level requirement, decomposes it, and gates CI on the assessment. In the following, Listing~\ref{lst:flow} shows the workflow, and Listing~\ref{lst:test} shows the tier-3 step.

Running \texttt{pytest} produces evidence; \texttt{vnvspec assess} builds the trace graph, checks that \texttt{REQ-USER-001} is covered through \texttt{REQ-MOD-003}, and fails CI if it is not. \texttt{vnvspec export} then produces the report, the compliance matrix against the chosen standard, and the GSN argument, all from the same objects. Nothing in the flow required the team to leave Python and adopt another tool. The same flow is designed to admit an AI agent in the authoring loop. The decomposition steps, tier 1 to tier 2 to tier 3, are exactly the steps a coding agent performs implicitly when it turns a user request into code and tests. \textsc{VNVSpec} gives that process an explicit, checkable artifact: the agent proposes requirements and decompositions as \texttt{Requirement} objects, the GtWR checker and the human review them as data, and the trace graph records what the agent claimed.

\begin{figure}[h]
\begin{lstlisting}[language=Python, caption={From catalog and user requirement to CI verdict.}, label={lst:flow}]
from vnvspec import Requirement, Spec
from vnvspec.catalog.ml import pytorch_training

# Tier 1: user requirement + catalog baseline
user_req = Requirement(
    id="REQ-USER-001",
    statement="The service shall not return a "
              "diagnosis when input quality is "
              "outside the ODD.",
    verification_method="test",
    acceptance_criteria=["Out-of-ODD inputs "
        "yield an abstain response."],
)
spec = Spec(
    name="diagnosis-service",
    requirements=[user_req],
).extend(pytorch_training.reproducibility)

# Tier 2: decomposition with a metric
mod_req = Requirement(
    id="REQ-MOD-003",
    statement="The input gate shall reject "
              "images with blur metric above "
              "0.35 (variance of Laplacian).",
    verification_method="test",
    acceptance_criteria=["Rejection rate on "
        "the blurred eval set is 100%."],
)
\end{lstlisting}
\end{figure}

\begin{figure}
\begin{lstlisting}[language=Python, caption={Tier 3: an ordinary pytest test, linked by marker.}, label={lst:test}]
import pytest

@pytest.mark.vnvspec("REQ-MOD-003")
def test_blur_gate_rejects_blurred_inputs(
        gate, blurred_eval_set):
    for img in blurred_eval_set:
        assert gate(img).action == "abstain"
\end{lstlisting}
\end{figure}

\subsection{Choosing an Evidence Route: pytest, Jest, and CROWN}
\label{sec:routes}

The evidence routes serve different situations, depending on where the test that ultimately verifies the requirement naturally lives. We illustrate when each route applies.

\textbf{Route 1: pytest.} If the decomposed requirement is implemented and checked in Python, the tier-3 test is an ordinary pytest test linked by a marker, as in Listing~\ref{lst:test}. This is the simplest route: evidence is captured during the test run the team already executes, and the marker itself serves as the trace link, so there is no separate export or conversion step that can fall out of date. It is the appropriate default whenever the property can be checked against concrete cases from within the Python test suite, whether the subject is a data-processing function, a service endpoint, or a trained model.

\textbf{Route 2: Jest through JUnit XML.} High-level requirements do not respect language boundaries. Consider the user requirement \texttt{REQ-USER-002}, ``the clinician interface shall display the abstain state whenever the service abstains,'' which decomposes to the React front-end that the team tests with Jest \cite{jest}. Rather than porting these tests to Python, the team keeps its native runner and transfers only the results, using JUnit XML, a test-result file format that most test runners can emit. The Jest test names the requirement in its title, and a three-line \texttt{jest-junit} configuration function turns that reference into a \texttt{vnvspec} property in the emitted JUnit XML (Listing~\ref{lst:jest}). The ingester then converts the XML file into requirement-linked evidence (Listing~\ref{lst:crown}, first part); skipped tests become inconclusive evidence, and references to unknown requirement IDs produce warnings rather than being silently dropped. Nothing in the JavaScript toolchain imports or depends on \textsc{VNVSpec}, and the same path serves any runner that emits JUnit XML, including those in the Java and C\texttt{++} ecosystems.

\begin{figure}
\begin{lstlisting}[language=JavaScript, caption={A Jest test verifying a high-level UI requirement. }, label={lst:jest}]
// ResultView.test.jsx (Jest)
test("abstain banner shown [REQ-USER-002]",
    () => {
  render(<ResultView result={abstain()} />);
  expect(screen.getByRole("alert"))
    .toHaveTextContent("No diagnosis: input "
      + "outside supported domain");
});

// junitProperties.js (jest-junit hook):
// title reference -> <property name="vnvspec">
module.exports = (tc) => {
  const m = /\[(REQ-[A-Z0-9-]+)\]/
    .exec(tc.title);
  return m ? { vnvspec: m[1] } : {};
};
\end{lstlisting}
\end{figure}

\textbf{Route 3: CROWN (for deep learning models).} Modules that embed machine learning models often carry requirements that must hold for \emph{every} input in a continuous range, not just for the examples in a test set. Consider \texttt{REQ-MOD-004}: ``For every image in the blurred certification set and any perturbation $\delta$ with $\|\delta\|_\infty \le 8/255$, the blur score shall remain above the rejection threshold 0.35.'' A pytest test can sample perturbations, but no finite number of samples can establish a property that must hold across an entire continuous set, so example-based evidence for this requirement is inconclusive by nature. Neural network verification methods address exactly this situation. CROWN \cite{zhang2018crown} computes guaranteed lower and upper bounds on a network's outputs that hold for every input within a specified perturbation range, and auto\_LiRPA \cite{xu2020autolirpa} provides an open-source implementation. Listing~\ref{lst:crown} wraps the certification run in an \texttt{EvidenceCollector} and records the outcome with \texttt{kind="formal\_proof"}, attaching the perturbation budget, the method, and the worst certified bound.

\begin{figure}
\begin{lstlisting}[language=Python, caption={Certifying output bounds with CROWN and recording evidence.}, label={lst:crown}]
import torch
from auto_LiRPA import (BoundedModule,
                        BoundedTensor)
from auto_LiRPA.perturbations import (
    PerturbationLpNorm)
from vnvspec import EvidenceCollector

with EvidenceCollector(spec) as c:
    # Route 2: ingest the Jest run (JUnit XML)
    c.from_pytest_junit("jest-junit.xml")

    # Route 3: CROWN bounds on the blur score
    bm = BoundedModule(blur_net,
                       torch.empty_like(xs))
    xb = BoundedTensor(xs, PerturbationLpNorm(
        norm=float("inf"), eps=8 / 255))
    lb, _ = bm.compute_bounds(
        x=(xb,), method="CROWN")
    certified = lb > 0.35   # per-image guarantee

    c.record(
        "REQ-MOD-004",
        "pass" if bool(certified.all())
        else "inconclusive",
        kind="formal_proof",
        message=f"{int(certified.sum())}/"
                f"{len(xs)} images certified",
        method="CROWN", eps=8 / 255,
        worst_lower_bound=float(lb.min()),
    )
report = c.build_report()
\end{lstlisting}
\end{figure}

The mapping from certification outcome to verdict illustrates the importance of strict verdict policydd. If the certified lower bound stays above the threshold for every image, the requirement receives a \texttt{pass} verdict that holds over the entire perturbation range, a guarantee that no finite test suite can provide. If CROWN fails to certify an image, this is not a counterexample: the bounds are guaranteed to be correct but are not always tight, so a loose bound only means that certification did not succeed. The evidence is therefore recorded as \texttt{inconclusive}, which keeps the requirement visibly open rather than silently passing. Only a concrete violating input, found for example by a falsification method, justifies a \texttt{fail} verdict. In the final report, the formal-proof evidence appears alongside the pytest and Jest evidence under the same requirement graph, and the compliance matrix reports each item with its evidence kind, so an auditor can distinguish a sampled check from a formal certificate at a glance.

\subsection{Experimental Evaluation}
\label{sec:eval}

We evaluate the framework along four questions. \textbf{RQ1}: Is the specification language expressive enough for a nontrivial system, and does continuous self-assessment catch real defects? \textbf{RQ2}: How do the core operations scale with specification size? \textbf{RQ3}: What overhead do the integrations impose on existing test workflows? \textbf{RQ4}: How reliably does the quality checker detect seeded requirement defects?

\subsubsection{Setup}
All measurements use the released packages, \texttt{vnvspec} 0.3.2 (from PyPI) and \texttt{pytest-vnvspec} 0.2.0, on deliberately CI-class hardware: a 2-core Intel Xeon at 2.80\,GHz with 8\,GB of RAM, Python 3.11, and pytest 9.1 on Linux. This choice is intentional, since the framework's primary execution environment is a CI runner, not a workstation. We report medians of five runs (three for the two largest configurations). The benchmark scripts are part of the replication package.

\subsubsection{RQ1: Self-Application}
A specification framework should be able to specify itself. Since version 0.1.0, \textsc{VNVSpec} has been developed against its own specification, \texttt{self-spec.yaml}, kept in the repository and assessed in CI on every commit, under a stated policy that a change that breaks self-assessment is rolled back, not the specification. The self-spec currently holds 36 requirements (12 blocking, 17 high, 7 medium priority) covering the core model contracts, the trace layer, the quality checker, the registries and exporters, the error hierarchy, and meta-level requirements about backward compatibility and packaging. Each requirement is verified by the ordinary test suite, 449 tests under a 95\% branch-coverage gate, and the most recent released assessment reports passing evidence for every requirement with none inconclusive. The self-spec caught real regressions during development: the backward-compatibility requirement turned an implicit norm into a released contract, and the dependency-declaration requirement was added after a CLI import error on clean installs; its automated check has prevented recurrence.

The measured cost of this gate is small. End-to-end, \texttt{vnvspec validate} on the self-spec takes 0.54\,s of wall-clock time including interpreter startup. The component costs are smaller still: loading the YAML spec takes 31\,ms; quality-checking all 36 requirements takes 1.8\,ms and reports 62 warnings and zero errors; building the trace graph takes 0.2\,ms; and the exports take between 0.1\,ms (GSN) and 8.1\,ms (XLSX compliance matrix), with gap analysis against the 29-clause EU AI Act registry at 0.3\,ms.

\subsubsection{RQ2: Scalability}
Table~\ref{tab:scal} reports the median runtime of each core operation on synthetic specifications of 10 to 10,000 requirements. All operations scale approximately linearly. YAML serialization and parsing dominate at every size; the framework's own logic is comparatively cheap, with quality checking at 0.35\,s and graph construction at 0.12\,s even for 10,000 requirements. A 1,000-requirement specification, larger than any specification we have encountered in practice, round-trips through YAML in under 1.4\,s and is quality-checked in 29\,ms, which supports the claim runnable on every commit.

\begin{table}[t]
\caption{Median runtime (seconds) vs. specification size, on 2-core CI-class hardware. QC = quality check of all requirements.}
\label{tab:scal}
\centering
\footnotesize
\setlength{\tabcolsep}{10pt}
\begin{tabular}{@{}lrrrr@{}}
\toprule
\textbf{Operation} & \textbf{10} & \textbf{100} & \textbf{1{,}000} & \textbf{10{,}000} \\
\midrule
Construct + validate & $<$0.001 & $<$0.001 & $<$0.001 & 0.002 \\
Serialize to YAML & 0.005 & 0.048 & 0.510 & 5.42 \\
Load from YAML & 0.009 & 0.087 & 0.868 & 10.2 \\
QC (all requirements) & $<$0.001 & 0.003 & 0.029 & 0.352 \\
Build trace graph & $<$0.001 & $<$0.001 & 0.004 & 0.118 \\
Verdict roll-up & $<$0.001 & $<$0.001 & $<$0.001 & 0.002 \\
\bottomrule
\end{tabular}
\end{table}

\subsubsection{RQ3: Integration Overhead}
Two experiments measure what the bridge costs the developer. First, we ran a suite of 400 no-op pytest tests, each marked with a distinct requirement ID, against a 400-requirement specification, with and without the plugin enabled. The suite takes 0.90\,s without the plugin and 1.23\,s with it, including spec loading, marker validation, evidence capture for all 400 tests, and writing the JSON report: an overhead of roughly 0.8\,ms per test. Because the tests themselves do nothing, this is the worst case in relative terms; against any real test body the absolute per-test cost is what matters. Second, the JUnit XML ingester processes a 5,000-case result file, with embedded requirement properties and a 4\% failure rate, in 0.46\,s (about 11,000 cases per second), recovering all 4,800 pass and 200 fail verdicts correctly.

\subsubsection{RQ4: Quality Checker Under Seeded Defects}
We mutated each of the 36 self-spec requirement statements with four defect-seeding operators aligned with the GtWR rule classes, and counted a defect as detected when the mutant produces strictly more error- or warning-level findings than its unmutated original. Table~\ref{tab:mutation} shows the results. The checker detects appended escape clauses and removed acceptance criteria perfectly (36/36 each). Injected vague terms are detected in 17 of 36 cases; the misses occur when the original statement already triggers the same lexical rule, so the mutation adds no new finding under our strict counting. Appended second conditions evade detection entirely (0/36): the current conjunction rule does not flag the added ``shall'' clause. We report these negative results deliberately, and return to them in Section~\ref{sec:disc}. Two complementary observations: quality-checking the shipped artifacts yields zero error-level findings for both the self-spec (62 warnings) and all 116 catalog requirements (186 warnings), which reflects the intended two-tier use of severities, with errors gating and warnings advising.

\begin{table}[t]
\caption{Seeded-defect detection by the GtWR quality checker}
\label{tab:mutation}
\centering
\footnotesize
\setlength{\tabcolsep}{10pt}
\begin{tabular}{@{}llc@{}}
\toprule
\textbf{Operator} & \textbf{Seeded defect} & \textbf{Detected} \\
\midrule
Escape clause & ``, where possible and as appropriate'' & 36/36 \\
Missing criteria & acceptance criteria removed & 36/36 \\
Vague terms & ``fast and user-friendly'' appended & 17/36 \\
Packed conditions & second ``shall'' clause appended & 0/36 \\
\bottomrule
\end{tabular}
\end{table}

\section{Discussion}
\label{sec:disc}

The experimental results bear on the two claimed contributions in different ways, and we discuss each in turn.

For the first contribution, the typed specification language, RQ1 establishes expressiveness in the only way available short of a user study: the language specifies a nontrivial multi-package system, including meta-level requirements about backward compatibility and packaging that most requirement notations cannot express as checkable objects, and the resulting gate caught real regressions. The RQ4 results qualify the quality-checker component of this contribution honestly. Lexically visible defect classes are caught reliably: escape clauses and missing verification criteria, arguably the two defects with the highest downstream cost, are detected at 36/36. Structurally visible defects are not: a packed second condition passes the current conjunction rule unflagged, and vague-term injection is masked whenever the original already carries a finding of the same class. The checker should therefore be understood as the cheap first filter the design intends, one that raises the floor at authoring time, and not as a substitute for requirement review. The severity distribution across the shipped artifacts (zero errors, hundreds of advisory warnings over 152 requirements) shows the two-tier gate operating as designed in practice.

For the second contribution, the bridge to existing test tooling, RQ2 and RQ3 turn the phrase ``cheap enough to run on every commit'' from a design goal into a measurement. The full self-assessment gate costs 0.54\,s end-to-end; the plugin adds 0.8\,ms to each test of a suite that already exists; ingesting five thousand externally produced results costs half a second; and every core operation scales linearly to specification sizes an order of magnitude beyond any we have observed in practice, with the dominant cost being YAML parsing rather than any framework logic. These are the costs that determine whether traceability survives contact with a real CI pipeline, because a gate that developers can feel is a gate that gets disabled. The route experiments of Section~\ref{sec:routes} complement the numbers by showing that the same evidence model absorbs results from a Python runner, a JavaScript runner that has never heard of the framework, and a formal verifier, without modification to any of them.

\section{Limitations and Future Work}
\label{sec:discussion}

We identify three limitations along with future works.

\textbf{Evaluation validity:} All evaluation in this paper was conducted by the framework's own authors, on the framework itself, on a single machine, with seeded-defect operators of our own choosing. The measurements answer cost and scalability questions credibly, but they say nothing about authoring effort, defect-finding power relative to a baseline process, or audit-preparation time. The corresponding future work is a controlled study with external teams, measuring these three quantities against a documented baseline; the framework's report-diffing facility provides the instrumentation, and we plan to conduct the study with community developers.

\textbf{Depth of the quality checker:} The checker is heuristic, rule-based, and lexical, and RQ4 highlights its current state: it detects escape clauses and missing criteria perfectly but misses packed conditions entirely. More fundamentally, it inspects phrasing rather than correctness, so a precise, verifiable requirement can still be the wrong requirement; validation in the full sense still needs humans in the loop. The corresponding future work is a semantic tier for the checker, including conjunction analysis and language-model-assisted review, evaluated against the seeded-defect protocol.

\textbf{Ecosystem coverage:} The five catalogs encode best-practice judgments pinned to framework versions and will drift as those frameworks evolve; whether community contribution under the published inclusion gate scales is an open question. The largest future opportunity, however, is agentic development: an AI coding agent run inside the \textsc{VNVSpec} protocol must state the tier-1 requirements it believes the user has, pass them through the quality gate, decompose them with explicit metrics, and produce tests that trace back up, so that the human reviews requirements and verdicts rather than raw diffs, and an auditor receives the same compliance matrix a human team would produce. The current study provides the protocol and data model, and we deliberately left to future work to orchestrating agents against them.

\section{Conclusion}
\label{sec:conclusion}

Low-level test tooling and systems-engineering V\&V doctrine are both mature, but the connection between them has remained a manual and decaying artifact. This paper introduced \textsc{VNVSpec}, an open-source framework that makes that connection executable: typed requirements with quality checking, standards clauses and curated catalogs as importable data, an enforced traceability graph from user requirement to test evidence, evidence routes that span pytest, JUnit-emitting runners such as Jest, and formal bound-propagation tools such as CROWN, and role-specific exports generated from a single source of truth. The framework is self-hosting: it is developed against its own 36-requirement specification, verified by 449 tests under a 95\% branch-coverage gate, with a rollback policy on self-assessment failures. The experiments show that this discipline costs little, with the full self-assessment gate at 0.54\,s, per-test plugin overhead at 0.8\,ms, and linear scaling to specification sizes well beyond practical need, while the seeded-defect study marks precisely where the quality checker's rule-based approach ends and future semantic checking must begin. If AI systems and AI-written software are to be certified, the requirement-to-evidence chain has to become a first-class engineering object, and our work provides a concrete proposal to achieve it by combining software and systems engineering approaches.

\bibliographystyle{IEEEtran}
\bibliography{refs}

\end{document}